# COVID-19 Submodel for the NC MInD ABM

# Overview, Design Concepts, and Details Protocol

Kasey Jones[1], Alexander Preiss[1], Emily Hadley[1], Sarah Rhea[2], Caroline Kery[1], Marie C. D. Stoner[1], Joëlla W. Adams[1]

[1]RTI International, Research Triangle Park, North Carolina
[2]North Carolina State University, Department of Veterinary Medicine, Raleigh, North Carolina

## Abstract

This Overview, Design Concepts, and Details (ODD) Protocol is an ODD extension to an agent-based model (ABM) framework built for the North Carolina Modeling Infectious Diseases Program (NC MInD). The model, NC MInD ABM, can be used as a base model for various infectious disease simulations. In this document, we describe a submodel specifically designed to simulate COVID-19 cases and hospitalizations within North Carolina. We describe in detail how each piece of the submodel works and how it was created. We do not discuss specific simulation scenarios or results. This information is reserved for papers related to each use case of the submodel.

**Introduction**

This Overview, Design Concepts, and Details (ODD) Protocol contains only information related to the COVID-19 submodel that has been appended to RTI's geospatially explicit agent-based model of North Carolina's healthcare facilities. The ODD for the base model, further referred to as the location model, is kept separate, as it is used as the base model for different submodels.[1] For information on the underlying location movement and agent information, please see the original ODD. For each section below, only information related to the COVID-19 submodel will be included.

**1. Purpose and Patterns**

The purpose of this ABM is to simulate agent movement to and from healthcare facilities during the COVID-19 pandemic. There are several applications of this COVID-19 submodel. Example use cases are: (1) simulate the impact of the COVID-19 pandemic on hospitalizations and estimate future hospital capacity levels; (2) estimate transmission risk in nursing homes under different scenarios; and (3) provide external within-hospital models agent-specific admission information. This is an adaptation of our original COVID-19 ABM that was used to forecast COVID-19 hospitalizations in North Carolina throughout 2020.[2] The COVID-19 submodel has since been refactored to support multiple different use cases.

*ABM Patterns*

To validate the model, we evaluate the ABM by its ability to reproduce patterns related to SARS-CoV-2 case count infection severity levels, COVID-19 related hospitalizations, and additional purpose-specific patterns. The example patterns below were calculated using one model run of 30 days.

**Pattern 1: SARS-CoV-2 infections.** We compare the daily number of forecasted SARS-CoV-2 infections by NC county to the number of SARS-CoV-2 infections produced in the model throughout the 30-day model run. Figure 1 shows this pattern for one county.

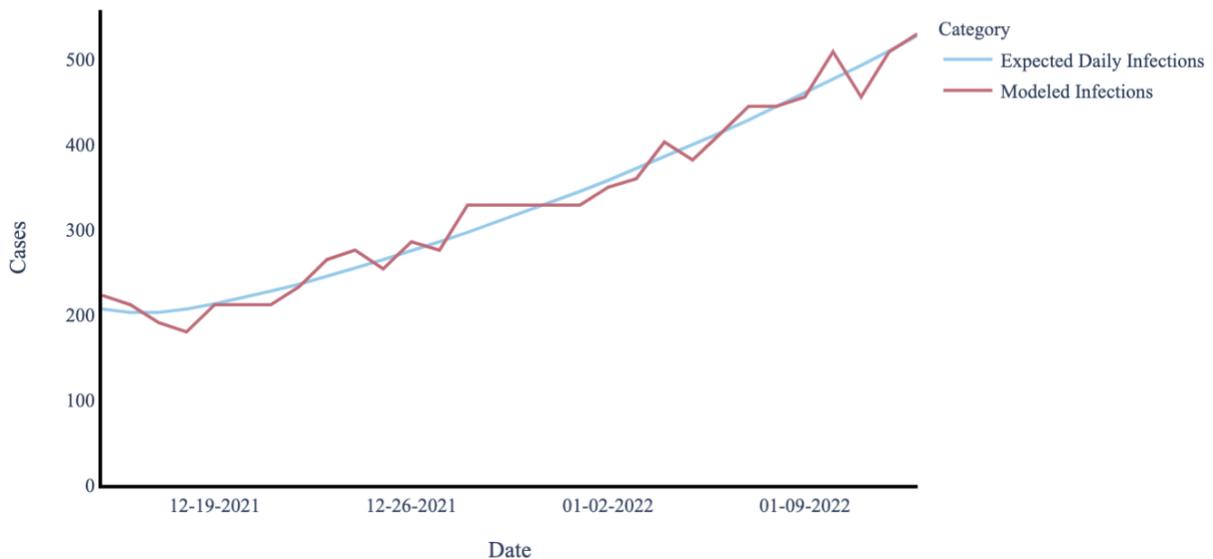

Figure 1. Example output of modeled cases vs. expected cases for a 30-day run

**Pattern 2: COVID-19 Outcomes & Case Counts.** We compare the proportion of COVID-19 case outcomes that occur in the model, to the expected outcomes based on input parameters and known values. We specifically look at hospitalizations (Table 1) but have created a suite of additional unit tests that check additional outcomes. We test the following outcomes:

- Proportion of cases that get reported
- Proportion of cases that are post vaccination
- Modeled cases by age and vaccination match expected case counts
- Comparison of case outcomes by vaccination and reported status (Table 1)
- Proportion of vaccinated hospitalizations match reported values
- Proportion of vaccinated ICU hospitalizations match reported values

- Proportion of COVID-19 cases resulting in hospitalizations match reported values

Table 1 shows example output of the modeled cases by vaccination status, age group, and case outcome. The full table is available on the project repository. For this pattern, we are comparing the modeled outcome proportion to the target proportion.

*Table 1: Example COVID-19 Case Outcomes by Vaccination Status and Age for Reported Cases*

| Vaccination Status | Age | COVID State | Modeled Cases | Modeled Proportion | Target Proportion |
|---|---|---|---|---|---|
| Not Vaccinated | 0 | Asymptomatic | 325 | 0.051 | 0.050 |
| Not Vaccinated | 0 | Mild | 5,998 | 0.933 | 0.935 |
| Not Vaccinated | 0 | Severe | 83 | 0.013 | 0.012 |
| Not Vaccinated | 0 | Critical | 23 | 0.004 | 0.003 |
| Not Vaccinated | 1 | Asymptomatic | 69 | 0.042 | 0.05 |
| Not Vaccinated | 1 | Mild | 1,473 | 0.913 | 0.904 |
| Not Vaccinated | 1 | Severe | 52 | 0.032 | 0.037 |
| Not Vaccinated | 1 | Critical | 19 | 0.012 | 0.009 |

**Pattern 3: Nursing Home visitations.** We have created a thorough unit test suite that ensures parameters for nursing home visitations are matched. We check the model's ability to match total visitations, visitations by age group, and frequency of visitations for each visitor. Please review the nursing home (NH) visitation unit tests for evaluation of this pattern.

**Pattern 4: Healthcare worker attendance.** For each nursing home, we calculate the ratio of the average daily number of hours worked by healthcare workers (HCWs) in a model run to the target number of hours from Centers for Medicare & Medicaid Services (CMS) data.[3] We assert that the average ratio is very close to 1 and that the distribution of ratios also around 1. In one example run, the mean ratio was 1.002 and the standard deviation was 0.044. In addition to this integration test, we test various steps in the process of creating HCWs, assigning them to facilities where they will work and modeling their attendance at work. Please review the HCW assignment and HCW attendance unit tests for further detail.

## 2. Entities, State Variables, and Scales

The ABM has two types of entities—agents and locations. Please see the original ODD for their descriptions. In addition to agent states of the base model, agents have three additional state variables.

*Table 2. COVID-19-Specific Agent State Variables*

| Agent State | Description | Dynamic | Type | Range |
|---|---|---|---|---|
| COVID-19 | COVID-19 state: Susceptible (1); Asymptomatic (2); Mild/moderate (3); Severe (4); Critical (5); Recovered (6) | Yes | Integer | 1–6 |
| Vaccination | Boolean for vaccination or not | No | Boolean | 1, 2 |
| HCW | Boolean for HCW or not | No | Boolean | 1, 2 |

## 3. Process Overview and Scheduling

In addition to the life and location updates of the base ABM, several COVID-19 related updates are included. As in the base model, each COVID-19 update is put into the queue of actions to be performed. Before actions are executed, all actions are placed in a random order. See *Section 8: Submodels* for details on the specific COVID-19 actions. The ABM is implemented with a 1-day time step. There is no sense of daily time in the model. After model initialization, the ABM runs for 30 days. Alternate lengths can be used but changing input parameters.

## 4. Design Concepts

We have only described design concepts below if they differ from the original ODD. This section is still being written.

*Interaction*

Agents in nursing homes interact in the form of a nursing home visit. These visits are recorded, but they do not influence SARS-CoV-2 cases, as we do not model disease transmission.

## 5. Initialization

The COVID-19 submodel is currently parametrized to December 2021. As the ongoing pandemic progresses, parameters will need to be updated. Parameters linked to sources that change over time were taken from those sources as of the specified start date.

*Vaccinations*

Vaccination status (vaccinated, not vaccinated) is initiated based on input parameters for the proportion of individuals that are vaccinated by group (Table 3). This status does not change for the duration of the model run and no additional vaccinations are given. This decision is based on the short time horizon of the model (30 days).

*Table 3. Vaccination Rates by Agent Group*

| Agents | Description | Value |
|---|---|---|
| HCWs | All agents designated as HCWs | .80 [4] |
| NH Residents | Agents initialized in an NH | .87 [4] |
| Community (< 50) | All remaining agents (including agents in hospitals) | .47 [5] |
| Community (50 < 65) | All remaining agents (including agents in hospitals) | .74 [5] |
| Community (65+) | All remaining agents (including agents in hospitals) | .92 [5] |

HCWs and nursing home residents receive a flat rate of vaccination, regardless of their home county or age. All other agents (i.e., community agents) receive vaccination assignments based on their ages and home counties. The probability of being assigned a vaccination at model initiation is $p_{ac}$, where $a$ is an agent's age group and $c$ is the home county. This value is based on the county's vaccination rate for a specific age group, $CR_{ac}$, the state's vaccination rate for a specific age group $SR_a$, and the input parameter for that age group, $IP_a$.

$$p_{ac} = \frac{IP_a}{SR_{ac}} * CR_{ac}$$

*COVID-19 Hospitalizations*

The base model will initialize agents in all healthcare facilities according to the original ODD. Additional agents from the community are selected to start in hospitals based on input parameters specifying how many severe (non-ICU hospital bed) and critical (ICU hospital bed) COVID-19 hospitalizations the model should initialize. Example starting values are found in Table 4, although this can be changed to fit alternative initial scenarios.

*Table 4: Count of Hospitalizations on Model Initiation by COVID Status*

| Agents | Description | Value |
|---|---|---|
| Severe | Agents are in an acute (non-ICU) hospital bed on 12/15/21 | 1,194 [6] |
| Critical | Agents start in an ICU hospital bed on 12/15/21 | 417 [6] |

This information is not hospital-specific, and agents from the community are selected based on hospitalized COVID-19 cases by age for North Carolina (Table 5).[7]

*Table 5: Proportion of Hospitalized COVID-19 Cases by Age*

| Age | Percentage of Hospitalized Cases |
|---|---|
| 0–50 | 31% |
| 50 < 65 | 25% |
| 65+ | 44% |

The purpose of this initialization is to have COVID-19 hospitalizations equal to the input parameters. Agents selected for COVID-19–related hospitalization are assigned a *remaining length of stay (LOS) value.* The process for creating this value is described in the original ODD. We use a COVID-19–specific LOS distribution when creating the *remaining LOS* distribution.[8(p19)]

*Community Infection*

Agents in the community also have a chance of being given a mild/moderate, asymptomatic, or recovered COVID-19 status to start the model. The number of agents starting with a SARS-CoV-2 infection or recovered status is determined by the start date of the model, as well as the output of the Susceptible-Infectious-Exposed-Recovered-Susceptible (SEIRS) compartmental models. These models are described in **Section 8**. Each county will have an estimate for the proportion of infectious and recovered individuals on day 0 of the model. Random individuals from the community are selected to match the distribution of case severity described in **Section 8**. Agents are selected based on their age, using the age distribution for reported cases in North Carolina (Table 6).[7]

*Table 6. Proportion of Reported and Hospitalized Cases by Age*

| Age | Percent of Reported Cases |
|---|---|
| 0–50 | 70% |
| 50 < 65 | 18% |
| 65+ | 12% |

Since severe and critical agents are already initialized using known hospital case counts, and any new severe or critical patients would require a hospitalization, initial community cases cannot be assigned a severity of severe or critical.

*HCWs*

HCWs are also assigned at model initiation. Currently, HCWs are only assigned for nursing homes. The model includes four types of HCWs: single-site full-time employees, single-site part-time employees, multisite employees, and contract workers. Each HCW is assigned to a facility or facilities where they will work. Assignment differs by HCW type and is driven by several model parameters. Overall, the process works as follows:

1. For each nursing home, a target number of HCWs in each category is calculated based on model parameters and input data (CMS payroll-based journal and provider information datasets—see below for details).
2. For the three employee HCW types, the number of HCWs per county is calculated, and agents are randomly selected by county to become HCWs.
3. Agents are randomly selected to become contract workers at the state level. Because they work at multiple facilities and do not have a primary facility, it is more difficult to determine in which counties they should live. Therefore, we rely on population density to place contract workers in appropriate geographic locations.
4. Employees are randomly assigned to primary facilities within their counties of residence.
5. Multisite employees are randomly assigned to secondary facilities. Facilities in their counties of residence are prioritized, but some multisite employees must have secondary facilities in other counties due to the requirement that their primary and secondary facilities must be different. In these cases, facilities closest to their counties of residence are prioritized.
6. Contract workers are randomly assigned to facilities. Like multisite workers, they may work in facilities in different counties.

The HCW assignment process is defined to meet the following criteria:

- Each facility should be assigned the target number of each type of HCW.
- No agent should be listed as an HCW more than once, including as more than one type.
- Each single-site employee should be assigned to exactly one facility.
- Each multisite employee should be assigned to one primary and one secondary facility.
- Each contract worker should be assigned to the number of facilities specified in model parameters.
- No more than 10% of multisite workers should have a secondary site in a county more than 50 miles from their home county.
- No more than 5% of multisite workers should have a secondary site in a county more than 100 miles from their home county.
- No more than 40% of contract workers should have a secondary site in a county more than 50 miles from their home county.
- No more than 20% of contract workers should have a secondary site in a county more than 100 miles from their home county.
- No more than 5% of contract workers should have a secondary site in a county more than 200 miles from their home county.

## 6. Input Data

This ABM has several data files specific to COVID-19.

*COVID-19 Reported Cases (submodels/covid19/data/cases/covid19_cases.csv)*

This file contains the number of confirmed COVID-19 cases by county and by day.[9] Instructions for downloading and cleaning these data are found in the repo.

*COVID-19 Vaccinations (submodels/covid19/data/vaccinations/vaccinations_by_age.csv)*

This file contains the number of vaccinations by county and by age.[7] Instructions for downloading and cleaning these data are found in the repo.

*CMS Payroll-Based Journal Data (submodels/covid19/data/pbj/PBJ.csv)*

This file contains nursing home staffing data. It is generated from two public CMS datasets: Daily Nurse Staffing[3] and Daily Non-Nurse Staffing.[10] Instructions for downloading and cleaning these data are found in the repo.

**7. Submodel**

**COVID-19**

Each day several disease specific steps are taken. Any update (action) that occurs will be added to the ABM's list of actions and placed in a random order before execution. There are currently four different actions.

*Action 1: Recovery*

Any agents whose recovery date for COVID-19 is the current date of the model will have their COVID-19 state set to "recovered."

*Action 2: New COVID-19 Case*

The ABM will randomly choose individuals from the community who are susceptible to be given a COVID-19 infection. This choice is made using a weighted probability based on an agent's age, using the same distribution used for community infection initiation.

Infection severity (asymptomatic, mild/moderate, severe, or critical) is assigned according to an agent's age, vaccination status, and reported status (Table 7 & 8).

*Table 7. Severity Probability for Reported Cases (12.5% of Cases)*

| Severity | Description | Not Vaccinated | Vaccinated |
|---|---|---|---|
| Asymptomatic | No symptoms; no change in agent behavior | .05 [11] | .25 [11] |
| Mild/Moderate | Some symptoms; no hospitalization; agent behaviors may change | Varies | .65 |
| Severe | Hospitalization required | Calibrated to NC Data | Calibrated to NC Data |
| Critical | Hospitalization in ICU required | Calibrated to NC Data | Calibrated to NC Data |

*Table 8. Severity Probability for Nonreported Cases (87.5% of Cases)* **

| Severity | Description | Not Vaccinated | Vaccinated |
|---|---|---|---|
| Asymptomatic | See Table 5 | .25 | .5 |
| Mild/Moderate | See Table 5 | .75 | .5 |

* A case multiple of 8 is used to estimate infections using reported case counts. Therefore, 12.5% of cases are reported and 87.5% of cases are not. We assume only reported cases can go to a hospital. More information on this case multiplier is in the SEIR model discussion later in this section.

** There is a lack of literature for nonreported cases. We assume that vaccinated cases will be more asymptomatic because reported vaccinated cases are more asymptomatic.

If an agent is assigned an asymptomatic or mild/moderate case, their infection will last 7 days (expert input). They are assigned a recovery day in the model, and their COVID-19 state will be set to "recovered" when the model reaches this day. During this time, asymptomatic individuals will have no behavior changes, as they do not know that they are sick. Mild/moderate individuals may undergo behavioral changes. HCWs will go to work at a reduced rate. This reduction is set to 80% and is based on frequency of HCW testing, and the assumption that most HCWs would not go to work if they were sick. Nursing home visitation for mild/moderate individuals is reduced by 60%.[12]

If an agent is assigned a severe (non-ICU hospital bed) or critical (ICU hospital bed) case, they will go to a hospital. They are also blocked from going to work (for HCWs) or visiting a nursing home. An LOS is assigned to them based on past COVID-19 hospitalizations (Table 9).[8]

*Table 9. COVID-19–Related Hospitalization LOS Parameters*

| Parameter | Description | Value |
|---|---|---|
| LOS mean | Mean LOS: The average number of days that admitted COVID-19 agents spend in a hospital. Used in a truncated normal distribution for sampling agent LOS. | 3 |
| LOS std | Standard deviation of LOS: The standard deviation in number of days that admitted COVID-19 agents spend in a hospital. Used in a truncated normal distribution for sampling agent LOS. | 5 |
| LOS min | Minimum LOS: The minimum number of days that admitted COVID-19 agents spend in a hospital. Used in a truncated normal distribution for sampling agent LOS. | 1 |
| LOS max | Maximum LOS: The maximum number of days that admitted COVID-19 agents spend in a hospital. Used in a truncated normal distribution for sampling agent LOS. | 50 |

*Action 3: NH Visitation*

When an agent is moved to a nursing home, they are assigned visitors. The nursing home resident is assigned zero to three visitors, and each visitor is assigned a daily probability of visitation for that resident. Visitors are selected by age, based on their visitor number. We assume an agent's first visitor will likely be older and visit more often, while each additional visitor would get younger and visit less often (Table 10). These values are meant to represent the average and not match the visitation patterns of individuals.[13–15]

*Table 10. Nursing Home Visitation Parameters*

| Visitors | Probability | Daily Visitation Probability | Age Distribution of Visitor |
|---|---|---|---|
| 0 | 15% | 0 | |
| 1 | 45% | .50 (15 times a month) | < 50 (10%); 50 < 65 (20%); 65+ (70%) |
| 2 | 25% | .16 (5 times a month) | < 50 (20%); 50 < 65 (40%); 65+ (40%) |
| 3 | 15% | .03 (once a month) | < 50 (40%); 50 < 65 (40%); 65+ (20%) |

Each day, agents living in a nursing home have visitation simulated. Each visitor's probability of visitation is compared to a random number to see if a visit occurs. Visitation can be turned off completely if the COVID-19 model is being used for other purposes. For a visit to occur, several barriers must be passed:

1. The visitor must be randomly selected to visit that day.
2. The visitor must still be in the community.
3. The visitor must still be alive.
4. The visitor cannot have severe or critical symptoms.
5. Visitors with mild symptoms have a chance of choosing not to visit.
6. A nursing home may require that an individual show proof of vaccination.
7. Facilities may limit the number of visits allowed per person per week or month.

If all barriers are passed, a visit is recorded. We track the number of visits and the COVID-19 state of the visitor and the nursing home resident.

*Action 4: HCW Attendance*

During model initialization, some agents are randomly selected to be one of four types of HCW (see ***Section 6: Input Data*** for details). Each day, each HCW's attendance to work is simulated. First, a random draw is compared to the model parameter determining the probability that any given day is a workday for that worker. We model all worker attendance as full workdays at a single facility per day. Therefore, part-time workers have a lower probability that any given day is a workday. If the day is a workday for a HCW, they attempt to go to work. As with nursing home visitation, several conditions must be met for the HCW to attend work:

1. The HCW must still be in the community.
2. The HCW must still be alive.
3. The HCW cannot have severe or critical symptoms.
4. HCWs with mild symptoms have a chance of choosing not to go to work.

If these conditions are met, the HCW attends work on that day. For HCWs who are assigned to more than one facility (multisite employees, contract workers), a random draw is used to determine which facility they will attend on that day. The HCW's attendance and COVID-19 status is recorded.

*COVID-19 Case Counts*

The output of the SEIRS models provides forecasted cases to be created by the model for each county and each day of the model run. This forecasted number of cases is then inflated using an input parameter to estimate infections. This parameter represents the underreporting of infections. Infections are further inflated to generate COVID-19 exposures for the model to create. The inflation of infections to exposures reflects the effectiveness and current level of vaccinations given the model's vaccination parameters. Vaccinated agents have a chance of being immune to COVID-19 based on the current estimation for vaccine effectiveness. This value is currently set to 24%. The value was estimated based on estimates of individuals in North Carolina who have received a booster shot (37% of the vaccinated population as of 1/5/2022) mixed with the effectiveness of the booster shot against the current COVID-19 variant (estimated to be around 50%). When an agent is selected to be exposed to COVID-19, but they are immune, the exposure is "blocked" and ignored in the model. Exposures that are not blocked become simulated COVID-19 cases. As an alternative to using SEIRS forecasts, models that simulate historical time periods can use actual COVID-19 case counts. This selection is set by the user within the input parameters.

To inflate infections for a county, $Inf_c$, to potential cases for the ABM, $PC_c$, we use the county vaccination rate, $Vacc_c$, and the overall vaccine effectiveness, $V_{eff}$. When combined, we get an estimate for the proportion of "blocked" cases. Estimated infections are inflated, so that after some cases are blocked, the number of estimated infections is created.

$$PC_c = \frac{Inf_c}{(1 - Vacc_c) + Vacc_c * (1 - V_{eff})}$$

The number of potential cases for each county, $PC_c$, will remain the same for each model run because the underlying county vaccination rate and underlying vaccine effectiveness do not change. However, the number of cases the model creates will change based on the scenario-specific county vaccination rate and vaccine effectiveness level.

**SEIRS Compartmental Models**

The ABM relies on daily SARS-CoV-2 infection projections. The model can be adapted to use any forecasts for county-level infections; however, we provide a submodel for creating these case projections. A SEIR model is a deterministic compartmental model used to simulate the spread of infectious disease. Because the ABM is being used to estimate infections during an ongoing pandemic, we have altered the SEIRS approach to account for available historical data. Using a single effective reproductive number, we run an individual SEIRS model for each NC county.

*Table 11. SEIRS Parameters*

| Parameter | Value | Description |
|---|---|---|
| Initial Case Multiplier | 10 (02/01/20–06/01/20) | Multiplier representing unreported SARS-CoV-2 infections [16–19] |
| Middle Case Multiplier | 4 (10/01/20–12/15/21) | Multiplier representing unreported SARS-CoV-2 infections [16–19] |
| Current Case Multiplier | 8 (12/15/21-present | Multiplier representing unreported SARS-CoV-2 infections; expert opinion |
| Length of Infection | 6 days | Average length of infection (expert opinion) |
| Length of Exposure | 5 days | Average number of days after exposure before someone has symptoms [20] |
| Immunity Length | 90 | Time after infection before moving from recovered to susceptible [21] |
| $R_e$ | Varies | Effective reproductive number |
| $R_0$ | 1.25 | Basic reproductive number (expert opinion based on case counts in NC 02/20–04/20) |

Before running the SEIRS model to create 30-day case forecasts, we estimate the value of each compartment. These values are calculated based on the number of reported cases through the start date of the ABM. Below we outline this process.

1. Case counts are smoothed using a 10-day rolling average. The smoothed cases are then scaled to ensure that the sum of the smoothed cases equals the sum of the reported cases.

2. The smooth case counts are multiplied by case multipliers based on the reporting date. This multiplier represents the estimated ratio of reported cases to total infections.

3. To estimate compartment $E_{ij}$: we divide the estimated infections for the next day by the population and model $\alpha$.

$$E_{ij} = \frac{Infections_{(i+i)j}}{population_j * \alpha}; \quad \alpha = \frac{1}{6}$$

4. To estimate compartment $I_{ij}$: we sum the previous 6 days of infections and divide by the county population. Note that infections and infectious ($I_{ij}$) are not the same.

$$I_{ij} = \sum_{i-6}^{i} \frac{Infections_{ij}}{population_j}$$

5. To estimate compartment $R_{ij}$: we calculate the rolling sum of compartment $I_{ij}$ that move to recovered and subtract out a proportion of the recovered compartment that would move back to susceptible. This proportion is calculated by shifting the proportion that went from $I$ to $R$ by the immunity length.

$$Total\ Recovered_{ij} = \sum_{d=1}^{d=i} I_{dj} * \frac{1}{6}$$

$$R_{ij} = Total\ Recovered_{ij} - Total\ Recovered_{(i-90)j}$$

6. To estimate compartment $S_{ij}$:

$$S_{ij} = 1 - E_{ij} - I_{Ij} - R_{ij}$$

Once the initial value of each compartment has been estimated for the start date, the compartment model is run for 30 days. Each county has its own model. The final estimates of the SEIRS models produce the number of estimated infections for each county and day for 30 days after the start date. This output is used to drive new SARS-CoV-2 infections for the ABM. Vaccination status is not currently used in the SEIRS model, although we have used SEIRS-V models at various points during this work.